\documentclass[12pt]{article}
\usepackage{mathpazo}
\usepackage[T1]{fontenc}
\usepackage[utf8]{inputenc}

\usepackage{xcolor}
\usepackage{amsmath}
\usepackage{bm}
\usepackage{relsize}
\usepackage{comment}
\usepackage{graphicx}
\usepackage[UKenglish]{babel}
\usepackage{microtype}                      
\usepackage{xcolor} 
\usepackage{hyperref}                   	
\usepackage{lipsum}
\usepackage{fancyhdr}
\usepackage{picture} 
\hypersetup{								
	colorlinks=true,      		            
	linkcolor=blue,                        	
	filecolor=blue,                     	
	citecolor=blue,    	        			
	urlcolor=blue,                 	   		
	pdffitwindow=false,               	 	
	pdfstartview={FitH},               		
	pdfauthor={Huw Price},           		
}

\usepackage{braket}
\usepackage{physics}
\usepackage{todonotes}
\usepackage{enumitem}

\newcommand{\V}{$\lor$}

\newcommand{\Vi}{{$\lor_{1}$}}
\newcommand{\Vii}{{$\lor_{2}$}}
\newcommand{\Valt}{\textsc{Big}-{$\lor$}}

\usepackage{tikz}
\usetikzlibrary{shapes,decorations,arrows,calc,arrows.meta,fit,positioning}
\tikzset{
    -Latex,auto,node distance =1 cm and 1 cm,semithick,
    state/.style ={ellipse, draw, minimum width = 0.7 cm},
    point/.style = {circle, draw, inner sep=0.04cm,fill,node contents={}},
    bidirected/.style={Latex-Latex,dashed},
    el/.style = {inner sep=2pt, align=left, sloped}
}

\begin{document}

\hypertarget{why-entanglement}{%
\title{Bell Correlations as Selection Artefacts}\label{bell}}
\author{Huw Price\thanks{Trinity College, Cambridge, UK; email \href{mailto:hp331@cam.ac.uk}{hp331@cam.ac.uk}.} {\ and} Ken Wharton\thanks{Department of Physics and Astronomy, San Jos\'{e} State University, San Jos\'{e}, CA 95192-0106, USA; email \href{mailto:kenneth.wharton@sjsu.edu}{kenneth.wharton@sjsu.edu}.}}
\date{\today}
\maketitle\thispagestyle{empty}

\begin{abstract}
\noindent We show that Bell correlations may arise as a special sort of selection artefact, produced by ordinary  control of the initial state of the experiments concerned. This accounts for nonlocality, without recourse to any direct spacelike causality or influence. The argument improves an earlier proposal in 
\cite{PriceWharton21b,PriceWharton22} in two main respects: (i) in demonstrating its application in a real Bell experiment; and (ii) in avoiding the need for a postulate of retrocausality. This version includes an Appendix, discussing the relation of the proposal to the conclusions of \cite{Wood15}.   \\

\end{abstract}

\section{Introduction}
\subsection{Overview}
We propose an explanation of the correlations characteristic of Bell experiments, showing how they may arise as a special sort of preselection artefact. This explanation accounts for nonlocality, without recourse to any direct spacelike causality or influence. If correct, the proposal offers a novel way to reconcile nonlocality with relativity.\footnote{The present paper strengthens our earlier   proposals in \cite{PriceWharton21b,PriceWharton22} 
 in two main respects: (i) in demonstrating its application in a real Bell experiment; and (ii) in avoiding the need for an explicit postulate of retrocausality.}  

We begin with a brief account of the landscape of discussions of the implications of Bell's Theorem, in order to explain where our proposal sits in relation to other approaches.
 \subsection{Orientation}
 In the discussions of issues arising from the work of Einstein, Podolsky, and Rosen and Schr\"odinger in the 1930s \cite{EPR,Sch35a,Sch35b}, and John Bell in the 1960s \cite{Bell64}, a key reference point is the Common Cause Principle (CCP). The following formulation of CCP will do for our purposes \cite{Hofer2013}: 
\begin{quote}
{The Common Cause Principle says that every correlation is either due to a direct
causal effect linking the correlated entities, or is brought about by a third factor,
a so-called common cause.}
\end{quote}
With reference to CCP, the relevant history goes like this. In 1935 EPR noted that correlations implied by what Schr\"odinger soon dubbed {`entanglement'} seemed to require explanation by common causes, not present within QM itself. EPR concluded that QM was incomplete, and Schr\"odinger concurred. The alternative -- the other option allowed by CCP -- was that measurement choices on one side of an experiment could influence results on the other, even though the two sides might be far apart, and spacelike separated. To EPR and Schr\"odinger, that sort of `nonlocal' influence seemed absurd; as Schr\"odinger put it, `that would be magic' \cite{Sch35a}.

In the 1960s, however, Bell proved that under plausible assumptions, the common cause option is untenable. That seems to leave us, as the above formulation of CCP puts it, with `a direct causal effect linking the correlated entities'; and hence with the kind of nonlocality that EPR and  Schr\"odinger believed to be absurd. As Bell saw, of course, this meant at least a \textit{prima facie} conflict with Relativity.

In broad-brush terms, we can classify responses to Bell's argument as follows. This taxonomy is not comprehensive or precise, but it will serve to locate our current proposal.
\begin{enumerate}
    \item \textbf{Accept nonlocality}, acknowledging its conflict with Relativity. Most such responses seek to mitigate the conflict, by arguing, for example, that Bell nonlocality is not the kind of full-blown (signalling) causation that would be in serious conflict with Relativity; or that the preferred frame required by nonlocality is not empirically detectable.\footnote{See \cite{Maud11} for a comprehensive defence of this option.}
    \item \textbf{Avoid nonlocality}, by arguing that Bell's result depends on an assumption of `Realism' or `Classicality', and rejecting this assumption.\footnote{See \cite{Maud14,Gomori23} for critical discussion of such views, and \cite{Wise17} for a more sympathetic treatment of the former version.}  
 
        \item \textbf{Render nonlocality compatible with Relativity}, by arguing that it is a partially \textit{retrocausal} process, acting via the past light cones of the two observers. This requires that we abandon Bell's assumption of \textit{Statistical Independence} (SI), so as to allow measurement choices to influence hidden variables (HVs) in their past.\footnote{See \cite{WhartonArgaman20} for a review of this approach. We discuss SI further in \S8.3 below. Note that this option requires that we make a choice about  the term `nonlocality'. There is a narrow use of the term, implying \textit{direct} spacelike influence, and a broad use, allowing the indirect influence proposed here. Those who prefer the narrow use will regard this option as another way of \textit{avoiding} nonlocality.}
    \item \textbf{Restore common causes} (and hence avoid nonlocality), by treating measurement settings as among the elements of reality influenced by factors in the common past of the experimenters. This option, called \textit{superdeterminism}, also requires violation of SI, and  hence is sometimes confused with option 3.\footnote{Some  authors, though not themselves guilty of this confusion, use the term `superdeterminism' for any view rejecting SI \cite{Hoss19}.} As this classification shows, however, it rests on a different choice between the two alternatives offered by CCP.
    \item \textbf{Seek to avoid the problem}, by arguing that the Bell correlations are not subject to CCP. Here there are at least two previous proposals. 
    \begin{enumerate}
        \item Arguing that the Bell correlations arise from the fact that our viewpoint as observers is always `perspectival', e.g., confined to one branch of a larger set of ‘many worlds’ (with no need
for nonlocality in the bigger picture).\footnote{As \cite{Myrvold21} put it, Bell's `entire analysis is predicated on the assumption that, of the potential outcomes of a given experiment, one and only one occurs, and hence that it makes sense to speak of \textit{the} outcome of an experiment.' }
    \item Rejecting CCP altogether, arguing that the lesson of the Bell correlations is simply that the world contains robust patterns of correlations not explicable in the two ways that CCP allows.\footnote{See \cite{vanF82} for a view of this kind. This view does not avoid nonlocality, of course. It merely declines to explain it as CCP requires.}
    \end{enumerate}  
\end{enumerate}
\subsection{Our approach}
In our own previous work we have explored option 3, seeking to make a case for retrocausality.\footnote{See \cite{Price96,PriceWharton15,WhartonArgaman20}.} Part of the case, of course, was  its evident potential to mitigate the consequences of nonlocality for Relativity. 

More recently \cite{PriceWharton21b}, we noted that retrocausality has the effect of introducing \textit{colliders} (i.e., variables influenced  by more than one contributing cause) into the causal model of Bell experiments, because the source event is influenced by measurement choices from both sides. This is interesting, in our view, because it suggests an avenue for explaining Bell correlations that bypasses CCP altogether. 

It is well known that `conditioning on a collider' -- i.e., selecting cases in which the collider variable takes a particular value -- can induce correlations between its independent causes. Such correlations do not call for explanation by one of the two routes permitted by CCP. They are merely `selection artefacts', as people say, and are not robust, in the sense of supporting difference-making counterfactuals. 

In recent work  \cite{PriceWharton21a}, we show that \textit{some} cases of entanglement -- namely, those due to delayed-choice entanglement-swapping -- are selection artefacts, in this sense.\footnote{This argument does not assume retrocausality; indeed,  the conclusion is especially straightforward if retrocausality is explicitly excluded.} As we argue there, it follows that EPR-Bell experiments that rely on such delayed-choice entanglement-swapping do not imply nonlocality, in the usual way. The correlations involved do not support the required counterfactuals.\footnote{The correlations concerned are only evident in certain post-selected ensembles, and a change in a measurement setting on one side of such an experiment might make a difference (say, via an intermediate quantum state) to the corresponding ensemble. So there is no need for it ever to make a difference to the outcome on the far side of the experiment. See \cite{PriceWharton21a} for details, and \cite{Guido21}, for similar results.}

Building on this work, we proposed that ordinary cases of entanglement might be explained in these terms, if we add two ingredients: retrocausality, and normal  
control of the initial conditions of an experiment \cite{PriceWharton22}. As we explain below, the latter factor turns out to account for the difference between the non-robust correlations in the delayed-choice entanglement-swapping case, and the robust, counterfactual-supporting Bell correlations in ordinary EPR-Bell experiments.

 The present piece improves our proposal 
 in two ways: first, by working through its application in a real Bell experiment; and second, importantly, by eliminating the need for a independent assumption of retrocausality. We show that we can get all the retrocausality the proposal needs by considering an imaginary case in which ordinary initial control is absent. (We compare this to the use of frictionless idealisations in mechanics.) In this imaginary case, causality is time-symmetric by definition. This gives us the colliders on which the proposal depends, while at the same time explaining why the retrocausality in question is lacking in the real world, in which ordinary initial control is not absent. 

 In terms of our taxonomy in \S1.2, our proposal is a hybrid. It agrees with 5(a) and 5(b) that the observed phenomena of Bell correlations do not call for the application of CCP, once properly understood, although not for the reasons that those options propose.\footnote{Our view doesn't postulate additional real phenomena to which we don't have access, as the Everett view does. And unlike 5(b), it doesn't claim that Bell experiments involve a \textit{new} kind of CCP-independent correlation -- on the contrary, we propose that Bell correlations turn out to fall under a \textit{familiar} exception to CCP.} Yet it agrees with option 3 that there is a robust counterfactual-supporting connection between the two wings of a Bell experiment, albeit one that is explained as a special sort of selection artefact, rather than as the kind of causal link mandated by CCP, once common causes are abandoned. As in the case of option 3, this connection is fully explained by connections within the light cones, so doesn't require any primitive spacelike connections, or preferred frame.

 \section[The Dortmund model]{The Dortmund model}

Let's begin with two versions of a familiar \V-shaped Bell experiment. The term `\V-shaped' alludes to the spacetime geometry of the experiments concerned, in a typical diagram with time on the \textit{y-}axis. 

In the first version (\Vi), a pair of entangled spin-1/2 
particles is produced with parallel spins in the plane of eventual measurement; we label this initial state $I_1$.\footnote{For example, if all measurements are constrained to occur in the $x-y$ plane, the state $(\ket{\uparrow\downarrow}+\ket{\downarrow\uparrow})/\sqrt{2}$ has the desired property.  This is distinct from the singlet state $(\ket{\uparrow\downarrow}-\ket{\downarrow\uparrow})/\sqrt{2}$, which will be used in the next example.} The particles are sent to two observers, Alice and Bob, who each perform a spin measurement at one of three angles in the specified measurement plane, arranged at 120\textdegree\ from each other. Let $\{a,b\}$ be the settings and $\{A,B\}$ be the outcomes, in the usual notation. The probabilities predicted by quantum theory are as follows:
\begin{itemize}
    \item[] When $a=b$, $P(A=B)=1$, $P(A\neq B)=0$\\
When $a\neq b$, $P(A=B)=0.25$, $P(A\neq B)=0.75$.
\end{itemize}
The expression $P(A=B)=0.25$ is shorthand for two equally probable cases, $A\!=\!B\!=\!0$ and $A\!=\!B\!=\!1$ (each, in this case, with probability $0.125$).  Similarly $P(A\neq B)=0.75$ is shorthand for the two outcome pairs $A=0,B=1$ and $A=1,B=0$, each with probability $0.375$. 

The second version (\Vii) of the experiment is the same, except that we begin with a pair of particles with antiparallel spins (the singlet state); we label this state $I_2$. This yields the complementary probabilities: 
\begin{itemize}
    \item[] When $a=b$, $P(A=B)=0$, $P(A\neq B)=1$\\
When $a\neq b$, $P(A=B)=0.75$, $P(A\neq B)=0.25$.
\end{itemize}
Let's also imagine an experiment \Valt\   that runs both \Vi\ and \Vii, and mixes the results, randomly and in equal proportions.\footnote{We can actually imagine several versions of such an experiment -- more on some of those variations below (\S8.2).}  We call it `big' because it combines two component experiments,  \Vi\ and \Vii.

It is easy to check that in \Valt, the probabilities of having originated in $I_1$ or $I_2$, given the final results, are given by the following expressions. For originating in $I_1$ we have:
\begin{itemize}
    \item[] When $a=b$,   $P(I_1|A=B)=1$, $P(I_1|A\neq B)=0$\\
When $a\neq b$,  
$P(I_1|A=B)=0.25$, $P(I_1|A\neq B)=0.75$. 
\end{itemize}
For originating in $I_2$ we have the complementary values: 
\begin{itemize}
    \item[] When $a=b$,   $P(I_2|A=B)=0$, $P(I_2|A\neq B)=1$\\
When $a\neq b$,  $P(I_2|A=B)=0.75$, $P(I_2|A\neq B)=0.25$.
\end{itemize}

\Valt\ does not exhibit Bell correlations in its results as a whole. The Bell correlations of the two sub-ensembles cancel out, so that the four variables in $\{a,b,A,B\}$ are pairwise independent of each other.  The Bell correlations reappear, of course, if we sort the results of \Valt\ into the two sub-ensembles: those originating from $I_1$ (case \Vi) and those originating from $I_2$ (case \Vii). 

These facts imply that  in \Valt\  there are probabilistic dependencies between the four variables in $\{a,b,A,B\}$ and a variable $\mathbf{I}$, representing the initial state (i.e., $I_1$ or $I_2$). Each of the variables in $\{a,b,A,B\}$ is probabilistically dependent on $\mathbf{I}$, conditional on the remaining three variables in $\{a,b,A,B\}$. We depict these dependencies in Figure~\ref{fig:MM3}. We have marked in green the setting variables, over which Alice and Bob have experimental control. We stress that the dashed bidirectional arrows represent \textit{probabilistic} dependencies, but not \textit{causal} dependencies. In particular, Alice's and Bob's choice of settings would \textit{not} normally be taken to influence the initial state variable $\mathbf{I}$, as in Figure~\ref{fig:MM31}.

\begin{figure}
\centering
\begin{tikzpicture}
    \node[state,fill=green] (a) at (0,0) {$a$};
    \node[state] (I) at (3.1,-2.5) {$\mathbf{I}$};
    \node[state] (A) at (1.2,0) {$A$};
    \node[state] (B) at (5,0) {$B$};
    \node[state,fill=green] (b) at (6.2,0) {$b$};

   \path[dashed,<->] (a) edge (I);
    \path[dashed,<->] (b) edge (I);
     \path[dashed,<->] (A) edge (I);
      \path[dashed,<->] (B) edge (I);

      \node[draw=blue,dotted,fit=(a) (A), inner sep=0.2cm] (machine) {};
  \node[draw=blue,dotted,fit=(b) (B), inner sep=0.2cm] (machine) {};
\end{tikzpicture}
\caption{Probabilistic dependencies in \Valt.} \label{fig:MM3}
\end{figure}
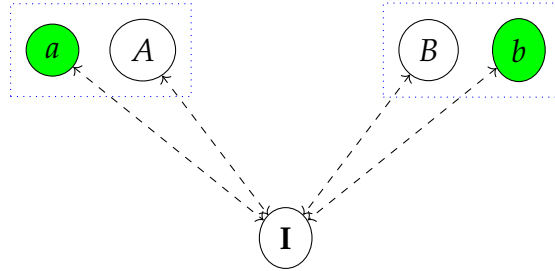

\begin{figure}
\centering
\begin{tikzpicture}
      \node[state,fill=green] (a) at (0,0) {$a$};
    \node[state] (I) at (3.1,-2.5) {$\mathbf{I}$};
    \node[state] (A) at (1.2,0) {$A$};
    \node[state] (B) at (5,0) {$B$};
    \node[state,fill=green] (b) at (6.2,0) {$b$};

   \path (a) edge (I);
    \path (b) edge (I);

      \node[draw=blue,dotted,fit=(a) (A), inner sep=0.2cm] (machine) {};
  \node[draw=blue,dotted,fit=(b) (B), inner sep=0.2cm] (machine) {};
\end{tikzpicture}
\caption{An \textit{unrealistic} causal  interpretation of \Valt} \label{fig:MM31}
\end{figure}
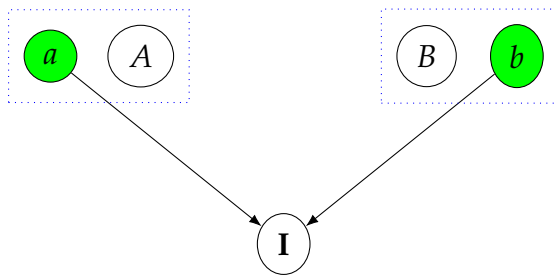

Let's think about what this means for the counterfactual judgements Alice makes, about what would have happened, had she made a different choice of the setting $a$. Since her choice does not influence $\mathbf{I}$, $\mathbf{I}$ would have had the same value, had she chosen differently. But we know that $B$ would have been different, in some cases -- that's Bell nonlocality at work.

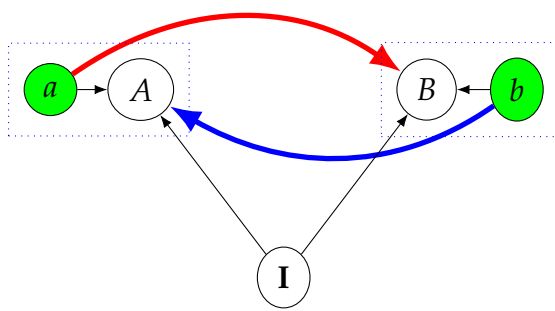
\begin{figure}
\centering
\begin{tikzpicture}
     \node[state,fill=green] (a) at (0,0) {$a$};
    \node[state] (A) at (1.2,0) {$A$};
    \node[state] (B) at (5,0) {$B$};
    \node[state,fill=green] (b) at (6.2,0) {$b$};
    \node[state] (I) at (3.1,-2.5) {$\mathbf{I}$};
 \draw[color=red,line width=2pt] (a) to [out=35,in=145] (B);
  \draw[color=blue,line width=2pt] (b) to [out=215,in=330] (A);

    \path (I) edge (A);
      \path (I) edge (B);
      \path (a) edge (A);
      \path (b) edge (B);

  \node[draw=blue,dotted,fit=(a) (A), inner sep=0.2cm] (machine) {};
  \node[draw=blue,dotted,fit=(b) (B), inner sep=0.2cm] (machine) {};
\end{tikzpicture}
\caption{\Valt\ with Bell nonlocality.} \label{fig:MM44}
\end{figure}

Figure~\ref{fig:MM44} shows the intuitive causal model for \Valt, or equivalently for each of the two sub-cases, \Vi\ and \Vii. The red and blue arrows represent the nonlocal influences implied by the existence of Bell correlations.  It is a contentious question exactly what sort of influences these are (e.g., whether they are really \textit{causal}).\footnote{For discussion of these issues, see \cite{Maud11,Wise17,Myrvold21}, and particularly \cite{Wood15}.}  We set aside that issue, and ask readers to interpret the blue and red arrows 
in terms of their own view of the matter.\footnote{We do require that the connection be counterfactual-supporting, so readers who are unconvinced about that may want to get off at this stop.} 

\section{What's wrong with Figure 2?} 
Why is Figure~\ref{fig:MM31} physically unrealistic? A number of answers suggest themselves: (i) causation works from past to future, not the reverse; (ii) the past is fixed, and therefore not amenable to influence by later choices; and (iii) as an initial variable, $\mathbf{I}$ is the kind of thing that an experimenter can control, or fix, before Alice and Bob choose their settings. 

The precise meaning of such answers, and their relation to one another, need not concern us here. But we note that for all of them, it is plausible that they are related in some way to the prevailing thermodynamic asymmetry in our universe (or our \textit{region} of the universe, if we don't want to exclude the possibility that it might be a local matter, as some have suggested). One thing that counts in favour of this suggestion is that in any of these three forms, we are trying to explain something \textit{time-asymmetric.} The thermodynamic asymmetry isn't the only possible physical basis for such time-asymmetries, but it is by far the most plausible one.\footnote{For discussion of this point, see, e.g.~\cite{Price96, Albert02,PriceWeslake10,Rovelli21}. Some authors argue that agency is a crucial ingredient here; but that, too, plausibly depends on the thermodynamic asymmetry.}

Again, we don't need to examine this argument in detail. What matters here is that we can \textit{imagine} the case in which the asymmetry that makes Figure~\ref{fig:MM31} physically unrealistic is absent. The suggested link with the thermodynamic asymmetry makes that relatively easy, because we seem to be able to imagine that it might be absent (or oriented differently in distant regions of our universe). 

For convenience, we adopt the label \textit{Initial Control} for the time-asymmetric feature of our world -- whatever it is -- that makes Figure~\ref{fig:MM31}  unrealistic. Our next step is to imagine a world  in which Initial Control is absent.

\section{Turning off Initial Control}

We want to imagine {turning off} Initial Control in \Valt. We need to imagine that the factors that normally enable control of initial conditions of an experiment are absent. It doesn't matter how we imagine this happening -- our argument doesn't require that it be physically realistic -- but turning off the Past Hypothesis seems a reliable, if drastic, way to achieve it!\footnote{By `Past Hypothesis' we mean the low entropy boundary condition apparently required in the early universe to explain the observed thermodynamic asymmetry, and all that rests on it. See \cite{Price96,Albert02,Price10,Rovelli21} for discussion.}

As a loose analogy, it may be helpful to compare turning off Initial Control to turning off friction. The frictionless case is wholly unrealistic, in many domains, but is nevertheless helpful to consider, not least to distinguish the effects of friction from other things.\footnote{It speaks in favour of this comparison that the time-asymmetric nature of friction is also linked to the thermodynamic asymmetry.}

Consider the effect of turning off Initial Control on the counterfactuals in \Valt. Normally, as we have seen, Alice will be entitled to treat $\mathbf{I}$ as fixed, in one direction or other, even if she doesn't know which.\footnote{We are setting aside for the moment  the question whether there might be \textit{actual} versions of \Valt\ in which we do not have control of the variable $\mathbf{I}$; see \S8.2.} 
This means she's entitled to say that {had} she chosen a different setting, $\mathbf{I}$ would have had the same value as it actually has; which means, as we said, that the difference would need to show up in Bob's outcome, in some cases. 
But with Initial Control turned off, Alice loses her entitlement to say that {had} she chosen a different setting, $\mathbf{I}$ would nevertheless have had the same value. That's what  turning off Initial Control means. It means not treating initial values (in this case, the value of the variable $\mathbf{I}$) as necessarily \textit{fixed,} for the purposes of one's causal modelling.\footnote{We don't mean that turning off Initial Control would give Alice and Bob this new kind of control of  $\mathbf{I}$ \textit{in any version of} \Valt\ \textit{whatsoever.} That seems unlikely. What's important is that it removes the usual barrier to embedding \Vi\ and \Vii\ within an imagined version of \Valt\ in which there is this kind of control.}

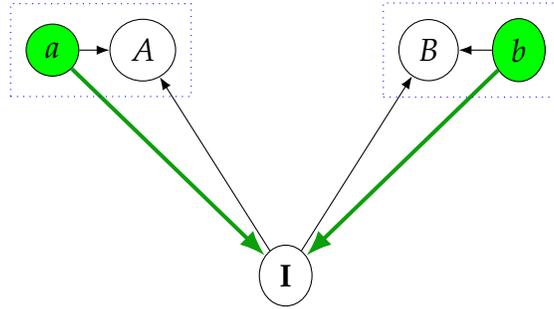
\begin{figure}
\centering
\begin{tikzpicture}
    \node[state,fill=green] (a) at (0,0) {$a$};
    \node[state] (I) at (3.1,-3) {$\mathbf{I}$};
    \node[state] (A) at (1.2,0) {$A$};
    \node[state] (B) at (5,0) {$B$};
    \node[state,fill=green] (b) at (6.2,0) {$b$};

   \path[color=green!60!black!95,line width=1.5pt] (a) edge (I);
    \path[color=green!60!black!95,line width=1.5pt] (b) edge (I);
      \path (a) edge (A);
      \path (b) edge (B);
      \path (I) edge (A);
      \path (I) edge (B);

      \node[draw=blue,dotted,fit=(a) (A), inner sep=0.2cm] (machine) {};
  \node[draw=blue,dotted,fit=(b) (B), inner sep=0.2cm] (machine) {};
\end{tikzpicture}
\caption{Without Initial Control -- $\mathbf{I}$ may become a collider.} \label{fig:MM32}
\end{figure}

Once Alice allows that in this imagined case -- with Initial Control turned off -- her choices might make a difference to   the value of $\mathbf{I}$, then she should model the case in the way shown in Figure~\ref{fig:MM32}. The green arrows represent the new possible dependencies, admitted by turning off Initial Control. The new feature of this case that interests us is that Figure~\ref{fig:MM32} treats the variable $\mathbf{I}$ as a \textit{collider,} in causal modelling terms.\footnote{Two notes about Figure~\ref{fig:MM32}. First, the black arrows should be regarded as legacies of the causal structure shown in Figure~\ref{fig:MM44}, the case in which Initial Control is not turned off. Other than adding the new green arrows, we set aside questions about the effect of turning off Initial Control on the dependencies shown in Figure~\ref{fig:MM44}. Second, we acknowledge that there are reasonable questions about the meaning of causal dependence in this imaginary case. We set those aside, too, simply noting that whatever account we use, it needs to be time-symmetric in this case -- that's the point.}

\section{Conditioning on a collider}
\subsection{General considerations}

A \textit{collider} is a variable with more than one direct cause, within a causal model. In other words, in the graphical format of directed acyclic graphs (DAGs), it is a node at which two or more arrows converge (hence the term `collider'). 

It is well known that conditioning on such a variable -- i.e., selecting the cases in which it takes a certain value -- may induce a correlation between its  causes, even if they are actually independent. As \cite[417]{Cole10} put it, `conditioning on the common effect imparts an association between two otherwise independent variables; we call this selection bias.'\footnote{Collider bias is also called {Berkson's bias,} after a Mayo Clinic physician and
statistician who noted it in the 1940s \cite{Berkson46}. But the point dates back at least to the Cambridge economist A C Pigou \cite{Pigou11}. (We are grateful to Jason Grossman and George Davey Smith here.)}

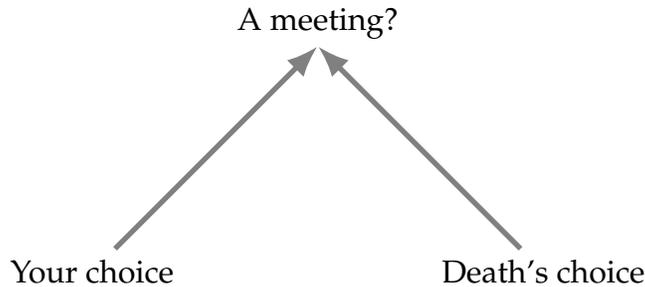
\begin{figure}
\centering
\begin{tikzpicture}
    \node (a) at (0,0) {Your choice};
    \node (F) at (3,3.3) {A meeting?};
    \node (b) at (6,0) {Death's choice};
\coordinate (coll) at (3,3); 
  \path[color=gray,rounded corners,line width=2pt] (a) edge (coll);
   \path[color=gray,rounded corners,line width=2pt] (b) edge (coll);

\end{tikzpicture}
\caption{A simple collider (`Death in Damascus')} \label{fig:M1}
\end{figure}

Here's a simple example, adapting the so-called `Death in Damascus' case, familiar in decision theory   \cite{Gibbard78}. Suppose that you and Death are each deciding where to travel tomorrow. You have the same two possible destinations -- in the usual version, Damascus and Aleppo. As in Figure~\ref{fig:M1}, your choice and Death's choice both influence an (aptly named) collider variable, which determines your survival. Let this variable take value $0$ if you and Death do not meet, and $1$ if you do. (We assume for simplicity that if the two of you choose the same destination, you will meet; and that this will be fatal, from your perspective.)

If we sample statistics for this kind of case, across a large population, they may suggest that people have an uncanny ability to evade Death, always choosing the opposite destination. If so, we've probably introduced selection bias, by interviewing survivors only. 

If you are a survivor, you might think to yourself, ``I'm a survivor, so if I had chosen the other destination, Death would also have made a different choice.'' You would be wrong. If you had chosen the other destination, you wouldn't have been a survivor. This illustrates an important fact. The correlations that result from conditioning on a collider do not support counterfactuals (in normal circumstances -- we'll come to an exception in a moment).

\subsection[Colliders in \Valt]{Colliders in \Valt}
Let's return to our imagined version of \Valt, with Initial Control turned off, depicted in Figure~\ref{fig:MM32}. 
As Alice reflects on this imagined case, these notions from causal modelling give her an obvious way of thinking about the Bell correlations in \Vi\ and \Vii. They look like selection artefacts, each resulting from conditioning on one of the two possible values of the collider variable $\mathbf{I}$. 

As we just saw, selection artefacts don't support counterfactuals, but in this imagined case they don't need to. In this case, Alice's choices make a difference to $\mathbf{I}$, and so it is a plus, not a minus, that they wouldn't support the counterfactuals associated with nonlocality. Alice can say, ``I thought my measurement choices were making a difference to the outcomes on Bob's side of the experiment, but that was a selection artefact. They were actually making a difference to the value of $\mathbf{I}$.''

The reader may wonder how this imagined case could possibly be relevant to the real world. In the real world, after all:
\begin{itemize}
\item Initial Control is \textit{not} turned off 
\item The causal model in Figure~\ref{fig:MM32} is \textit{un}realistic 
\item Bell correlations \textit{do} support counterfactuals.
\end{itemize}
 Remarkably, there's a way to solve all three problems, in one step. All we need to do is to turn Initial Control back on. But to explain why this does the trick, we need one more observation about colliders.
 
 \section{Constrained colliders}

In our previous work we have introduced the notion of a \textit{constrained} collider \cite{PriceWharton21b,PriceWharton22}. Intuitively, this is a restriction imposed from outside a causal model, biasing or completely specifying the value of the variable at the collider. 

In Death in Damascus, for example, we saw that your choice and Death's choice both influence a collider variable that takes value $0$ if you do not meet, and $1$ if you do. If Fate wants to ensure that your number's up, as it were, she \textit{constrains} this collider, setting its value to $1$. In effect, Fate imposes a future boundary condition, \textit{requiring} that the collider variable take value $1$.

This boundary condition makes a big difference to the counterfactuals. If it weren't for Fate's involvement, grieving relatives would be entitled to say, ``If only they had made the other choice, they would still be with us today!'' Once Fate constrains the collider, this is no longer true. If you'd made the other choice you would have met Death in the other place, instead. 

With Fate constraining the collider, in other words, there is a counter-factual-supporting connection between your movement and Death's. You control Death's movements, in effect.\footnote{This is the kind of thing that makes the Death in Damascus case interesting for decision theorists, of course; see \cite{PriceWeslake10} for discussion.}  This is what we term \textit{Connection across a Constrained Collider} ({CCC}) \cite{PriceWharton22}.

Collider constraint can come by degrees. Fate might be kind to you, and give you some small chance of eluding Death, at least tomorrow. We will be interested here in the full constraint version, in which case we'll say that the variable at the collider is \textit{locked.} 

A locked variable can only take one value -- that's the point. This means that it is no longer really a \textit{variable} at all, in the usual sense of a causal model, and can no longer be an effect of any of the remaining variables. We could put it like this: causation requires making a difference, and locking prevents making a difference. By locking the collider variable in the Death in Damascus case, Fate makes it the case that your choice makes no difference to whether you encounter Death. You no longer have any causal influence on the matter.

This may seem highly unrealistic, but there
is at least one place in physics where collider constraint has actually been proposed. It is the key to the so-called Horowitz-Maldecena hypothesis, concerning the  black hole information paradox. 
Horowitz and Maldacena describe the proposal as follows:  
\begin{quote}
In the process of black hole evaporation, particles are created in correlated pairs with one
falling into the black hole and the other radiated to infinity. The correlations remain even
when the particles are widely separated. The final state boundary condition at the black
hole singularity acts like a measurement that collapses the state into one associated with
the infalling matter. This transfers the information to the outgoing Hawking radiation in
a process similar to ``quantum teleportation''. \cite{HorowitzMaldacena04}
\end{quote}
The key difference from ordinary quantum teleportation is that the `final state boundary condition' imposes a particular result on the measurement concerned, thus eliminating the usual need for postselection. In our terminology, this amounts to constraining a collider at that point.

Discussing the Horowitz-Maldacena hypothesis recently, Malcolm Perry puts it like this:

\begin{quote}
{[}t{]}he interior of the black hole is therefore a strange place where
one's classical notions of causality \ldots{} are violated. This does
not matter as long as outside the black hole such pathologies do not
bother us. \cite[9]{Perry21}
\end{quote}
As we'll explain, our proposal is going to be that boundary conditions doing this job are
actually extremely common, if you know where to look. In the other
direction of time, they are just ordinary initial boundary conditions, and don't need black holes.

 \section{Turning Initial Control back on}

We noted two points about constrained colliders. Connection across a Constrained Collider (CCC) does support counterfactuals. And a \textit{locked} collider is removed from its causal model altogether, in the sense that because it is locked, it cannot be influenced by variables elsewhere.

This gives Alice a way of applying the lessons of the imaginary case of \Valt\ (without Initial Control) to the real case. Turning Initial Control back on -- adding it by hand to the imagined version of \Valt, as it were -- produces a model applicable to real-world  operational versions of \Vi\ and \Vii. Each of these experiments may be regarded as a locked case of the imagined version of \Valt. In \Vi, Initial Control locks the variable $\mathbf{I}$ to value $I_1$; in \Vii\ it locks it to value $I_2$. 

As just noted, locking the collider at $\mathbf{I}$ removes it from the causal model, thereby explaining why Figure~\ref{fig:MM31} and Figure~\ref{fig:MM32} are unrealistic. And it generates a connection across the collider that does support the counterfactual-supporting influences shown in red and blue in Figure~\ref{fig:MM44}. 

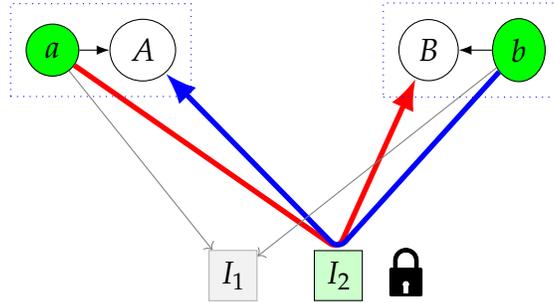
\begin{figure}
\centering
\begin{tikzpicture}
    \node[state,fill=green] (a) at (0,0) {$a$};
    \node[rectangle,draw=black,fill=green!20,inner sep=0.15cm] (F2) at (3.8,-3) {${I}_2$};
      \node[draw=gray!60,rectangle,fill=gray!10,inner sep=0.15cm] (F1) at (2.4,-3) {${I}_1$};
    \node[state] (A) at (1.2,0) {$A$};
    \node[state,fill=green] (b) at (6.2,0) {$b$};
    \node[state] (B) at (5,0) {$B$};
 \draw[color=red,rounded corners,line width=2pt] (a) -- (3.8,-2.65) -- (B);
  \draw[color=blue,rounded corners,line width=2pt] (b) -- (3.8,-2.65) -- (A);
    
       \path[color=gray,->] (a) edge (F1);
    \path[color=gray,->] (b) edge (F1);
    \path (a) edge (A);
      \path (b) edge (B);

  \node[draw=blue,dotted,fit=(a) (A), inner sep=0.2cm] (machine) {};
  \node[draw=blue,dotted,fit=(b) (B), inner sep=0.2cm] (machine) {};
   \node (myfirstpic) at (4.7,-2.95) {\includegraphics[width=18pt]{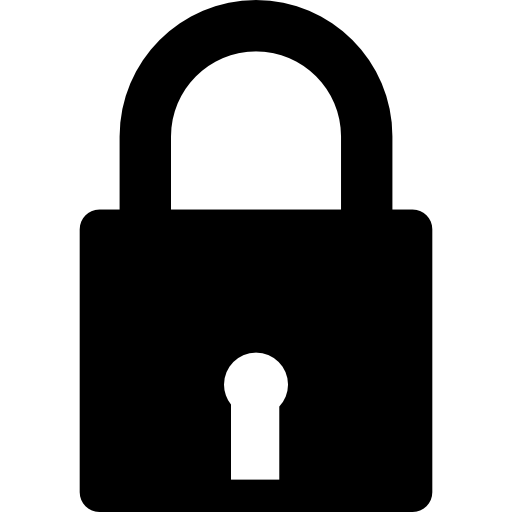}};
\end{tikzpicture}
\caption{\Valt\ with the dependencies induced by a locked node at ${I}_2$.} \label{fig:MM4}
\end{figure}

This proposal has a huge payoff, from Alice's point of view. It gives her a way of explaining the Bell correlations, without any relativity-challenging nonlocality. 
Referring to Figure~\ref{fig:MM4}, in other words, Alice can say that the Bell correlations in \Vii\ are explained by the fact that \Vii\ may be regarded as a product of a locked collider, the collider itself having the structure shown in Figure~\ref{fig:MM32}.

In Figure~\ref{fig:MM4} we again indicate by red and blue arrows the dependencies entailed by the Bell correlation. Now, however, it seems appropriate to represent these dependencies as linking the two sides of the experiment \textit{via} the locked node at ${I}_2$. If nothing else, this depiction serves to emphasise that there is no direct spacelike influence involved. The connection arises from the locked collider, which is in the overlap of the past light cones of Alice and Bob.

We stress again that   the retrocausality that Alice needs in  Figure~\ref{fig:MM32}, to explain Bell correlations in this way, is something that \textit{comes for free} when Initial Control is imagined turned off.
 It is an ingredient that becomes available anyway, to the extent that the argument needs it, when we consider the case in which Initial Control is absent.\footnote{Contrary to our own argument in \cite{PriceWharton22, PriceWharton23}, then, this recipe for explaining entanglement needs three ingredients, not four!}

Of course, Alice might have some other notion of causality in mind, such that turning off Initial Control is not sufficient for retrocausality \textit{in her sense.} For example, she might regard it as true by definition (according to what she means by `causality') that causation only works forward. That  doesn't matter for our argument. We don't require that the variable $\mathbf{I}$ be a collider {in Alice's sense of causation,} whatever that might be; but simply that it be a collider in the minimal sense that comes for free by turning off Initial Control.

We also emphasise that the argument does not require that Alice has retrocausal influence over the initial state of \textit{actual} experiments of the form of \Vi\ and \Vii. It simply requires that she not rule out such influence in the imaginary case in which Initial Control is missing -- which turns out to be true by definition, in the sense of causality that matters.  
 This is sufficient to make sense of a version of \Valt\ without Bell correlations, from which the Bell experiments \Vi\ and \Vii\ emerge by a kind of preselection: by fixing the value of the initial variable $\mathbf{I}$. In other words, it shows how Bell correlations in real experiments can be a selection artefact. 

\section{Discussion}

\subsection[The real world is a lot bigger than \Valt]{The real world is a lot bigger than \Valt}
In the argument above, \Valt\ is a stand-in for something much bigger. It takes a huge amount of Initial Control to prepare a version of \Valt, of course, even if we don't take the additional step of choosing between \Vi\ and \Vii. So we haven't really got rid of Initial Control -- we've just moved it a little further out. Still, we've provided a simple but realistic model on which more general versions can be built, and a proof of principle. Intuitively, enlarging the model will provide more, rather than fewer, ways in which Bell correlations can emerge as preselection artefacts.

\subsection{Physically realistic unlocked versions of \Valt?}
We have been assuming that any physically realistic version of \Valt\ would be the kind that mixes results from versions of \Vi\ and \Vii\ in which the initial state is indeed locked. But is this really the case, or are there realistic versions of \Valt\ in which $\mathbf{I}$ is not locked? We leave this as an open question,\footnote{The delayed choice framework seems a promising place to look.} 
 and recommend it among other things as a training exercise. As we said, our argument does not depend on a positive answer, but it does require the ability to imagine that the answer \textit{might} be positive. That's a step on the path to imagining the case in which there is no Initial Control.

\subsection[Two paths to retrocausality]{Two paths to retrocausality\footnote{Not to be confused with Adlam's  \cite{Adlam22} two \textit{roads} to retrocausality!}}
There are important differences between the approach taken above and typical retrocausal approaches to the explanation of Bell correlations.\footnote{See \cite{FriedrichEvans19,WhartonArgaman20,Adlam22} for recent surveys of the latter.} We noted one of these in \S1.2. In terms of our taxonomy there, typical retrocausal approaches fall into category 3. They seek to make Bell's nonlocality relativity-friendly, by representing it as a zigzag process, via past light cones. The present approach aims instead to remove Bell correlations from the scope of CCP altogether, by construing them as selection artefacts.

This difference is linked to another. So far as we are aware, all previous work in this area (including our own)  considers retrocausal influences on properties conventionally denoted by $\lambda$, understood to be the factors \textit{not} fixed by the initial preparation of the state $\psi$ of the quantum system concerned. 

As noted in \S1.2, retrocausality requires a violation of a key assumption required by Bell's Theorem, the principle called Statistical Independence (SI), or Measurement Independence.  
Expressed in terms of $\lambda$, Statistical Independence amounts to this:
\begin{description}
   \item [\normalfont(SI)]$P(\lambda|\alpha)=P(\lambda)$ (where $\alpha$ is a setting variable). 
\end{description}
Standard retrocausal models allow the possibility that $P(\lambda|\alpha)\neq P(\lambda)$.

In the usual approach, $\psi$ is an input variable, and is simply not assigned a probability. Hence it doesn't make sense to ask whether  
\begin{description}
   \item [\normalfont(SI$_\psi$)]$P(\psi|\alpha)=P(\psi)$. 
\end{description}
Turning off Initial Control changes this. $\psi$ can now be assigned a probability, if we wish, and turning off Initial Control can be then be regarded as an explicit rejection of the assumption SI$_\psi$. We remind readers that our argument requires this only in what we termed an imagined case, which can be physically unrealistic. The argument works even if in the actual world, $\psi$ is always subject to Initial Control. 

Why is this difference between these two approaches interesting? For at least two reasons, we think. First, the present approach doesn't require any \textit{actual} retrocausality, and thereby presents a much smaller target to would-be objectors. Second, the approach seems to be ontology-neutral, at least over a broad range,\footnote{The exceptions will include approaches that reject ontological questions altogether, in the quantum realm.} in the sense that whatever account is offered of the ontology of quantum state preparation -- whatever the ontological basis of $\psi$, in effect -- there will be a way of raising the possibility of turning off Initial Control \textit{in that ontology.}  

\subsection{Remembering the sub-operational level}

We have noted that the kind of retrocausality needed for our argument comes for free, when we imagine turning off Initial Control. Turning off Initial Control in \Valt\ \textit{just is} allowing the possibility that Alice's and Bob's settings choices might influence $\mathbf{I}$, in the sense that matters. 
But if Initial Control is turned off, so that Alice's and Bob's settings choices may make a difference to $\mathbf{I}$, how is this supposed to work? To what else, between $\mathbf{I}$ and the setting choices $a$ and $b$, can the latter also make a difference?

This question takes us to the sub-operational level, and we should expect any answers to be heavily dependent on choice of quantum ontology. We note the following general points. First, whatever goes into this sub-operational level, it will need to allow the kind of retrocausality, or future input dependence, that comes for free with giving up Initial Control.  It would be nonsense to imagine making a difference to $\mathbf{I}$ without making a difference to the proposed intermediate ontology. 

Second, the converse does not hold. That is, Initial Control of $\mathbf{I}$ \textit{need not} imply full Initial Control of sub-operational ontology between $\mathbf{I}$ and the settings. In the sense that has turned out to matter, retrocausality and lack of Initial Control go together. And in ordinary circumstances, we don't have reason to assume that we full control over HVs in the quantum realm -- quite the contrary. So the equation of lack of Initial Control with retrocausality, in the sense of the latter relevant to present discussions, seems to imply that retrocausality is always the default option in the HV case. This would be a striking reversal of fortune for the retrocausal approach, putting its would-be opponents on the back foot.

\section{Summary} 

We have shown that in a simple but realistic case, Bell correlations may be regarded as a special sort of preselection artefact, produced by ordinary control of the initial conditions of the relevant experiments. We conjecture that the reason this explanation has not been noticed is the sheer familiarity of the initial control concerned. To understand the role it plays we need to imagine it absent. This step has not previously been taken in this context,\footnote{It has been taken in a single-particle context, including by us; see, for example: \cite{Price12,PriceWharton15,LeiferPusey17}.} so far as we are aware.

When we do imagine the Initial Control-free regime, we bring into play the causal model of Figure~\ref{fig:MM32}, in which the initial state of \Valt\ is a retrocausal collider. In this regime, the Bell correlations in \Vi\ and \Vii\ can be explained as collider artefacts. Restoring Initial Control, taking us back to the real world, then \textit{constrains} the collider in question, a step known from other cases to convert mere collider artefacts into counterfactual-supporting connections, from one side of the constraint to the other. 

In effect, the initial state preparation that Initial Control allows thus `thwarts' the influence that Alice's and Bob's choices of settings would otherwise have on the same initial variable. Thwarted at that point, the difference due to different setting choices has to emerge somewhere else, and the result is Bell nonlocality.\footnote{We are grateful for comments from Michael Cuffaro. HP is also grateful to Heinrich P\"as and his group at TU Dortmund, for discussions that prompted some of the ideas presented here.}

\appendix
\section{Appendix: Relation to Wood \& Spekkens}

In an influential paper, Christopher Wood and Rob Spekkens  (WS)  discuss Bell correlations using the tools of the causal modelling and causal discovery literature \cite{Wood15}. In particular, they argue that \textit{any} causal model for Bell nonlocality must violate a principle known in that literature as \textit{Faithfulness.} Roughly, this principle states that the observed conditional independencies of a causal model should all reflect underlying causal independencies of the model in question. 

WS argue that causal models for Bell nonlocality need to reject Faithfulness to preserve \textit{no signalling.} The latter is an independence condition, and WS show that in order to respect it -- to  explain why Bell nonlocality cannot be used for signalling, despite the postulated causal structure -- any of a range of possible causal models for nonlocality will violate Faithfulness. In terminology familiar to physicists, they say, this amounts to a requirement for \textit{fine tuning.} 

We make three comments about the relation of the present proposal to WS's results and discussion. 
\subsection{Adding the present proposal to WS's range of options}

WS do not appear to have the present proposal on their radar, despite considering both retrocausality and the suggestion that the initial state $\psi$ might be added to the causal model. They discuss the latter suggestion in the context of the proposal that $\psi$ might be a common cause, and in the ArXiV version of their paper raise an objection which resonates with a key element of our proposal:
    \begin{quote}
        Within the framework of causal models, only a \textit{variable} can act as a common cause. If one takes the quantum
state of the pair of particles to be a common cause within this framework, then it is a trivial variable: it is fixed in the experiment to some particular state $\psi$, which is known to the experimentalist. (24)
    \end{quote}
In our terms, this characterises the usual case, in which Initial Control is present. It makes the same point as our observation that constraining a collider removes it from the causal model in question, because it is no longer a variable.

In the published version of their paper, WS add some further discussion:
  \begin{quote}
     If one takes the quantum
state of the pair of particles to be a common cause within this framework, then we must introduce a variable $\Psi$ 
that varies over all possible quantum states for the pair of particles and allow this to causally influence both \textit{A} and
\textit{B.} (23)
 \end{quote}
 All that has now been missed, by our lights, is the possibility that this new variable $\Psi$ might be a common \textit{effect,} rather than a common \textit{cause;} and hence that  the normal experimental control of $\Psi$ -- i.e., the ability to set its value to $\psi$ -- amounts to a form of preselection.  This would be conditioning on a collider, in causal modelling terms, and a \textit{constrained} collider, in our terminology.

 %
 

 \subsection{Two kinds of fine tuning}
 Our proposal, too, seems to rely on a form of fine tuning, albeit a very familiar one. As we noted, Initial Control appears to be a manifestation of the low entropy past of our universe (or region of the universe), which itself seems extremely fine tuned. 
 
 It is interesting to compare and contrast the role of fine tuning in the two approaches. In the models WS consider, fine tuning is needed to `turn off' signalling -- signalling is the default. In our proposal, it is needed to `turn on' CCC, by constraining the collider. The default is that there are \textit{no}  counterfactual-supporting connections outside the lightcones, not that there are stronger connections. This, as well as the fact that the low entropy past is needed for many other purposes, seems to make our appeal to fine tuning less `costly' than those identified by WS. 

    Still, it is not clear whether our approach avoids the difficulty that WS identify. We have argued that CCC is a robust, counterfactual-supporting connection. Whether we call it causation or not, it mimics causation in some respects. The fact that it depends on fine tuning in the form of Initial Control does not seem to immunise it against a requirement for \textit{further} fine tuning, to prevent its use for signalling. There is still a difference, in that the new question is more like `Why does nature not go further?' than `How does nature hide something that is there in any case?' But the lack of signalling still needs to be explained, apparently -- it is trivial to invent toy models in which CCC does support signalling. It would be interesting to examine existing proposals for explaining the kind of fine tuning that WS identify,\footnote{See, for example, \cite{Almada15, Evans15, Evans21}.} in the light of this new framing of the question.

    WS observe that Valentini's \cite{Val91a,Val91b} version of a deBroglie–Bohm model proposes to explain fine tuning, in their sense, in terms of an equilibrium distribution over a system's possible configurations. We note that in combination with an appeal to a low entropy boundary condition to explain Initial Control, this would lead to a familiar pattern of explanation. It is a common view that the familiar time-asymmetric thermodynamic character of our universe, and whatever depends on it, turns on a combination of two things:  (i) the \textit{Past Hypothesis} (PH), specifying a low entropy macrostate for the universe shortly after the Big Bang; and (ii) a \textit{Statistical Postulate} (SP), specifying a uniform probability distribution specified by the standard Lebesgue measure over the physically possible microstates that realize that initial macrostate.\footnote{This approach has been defended recently by Albert and Loewer, from whom we take this formulation -- see \cite{Loewer20} for a survey.} It would be a satisfyingly tidy result if the present approach allowed this framework to explain Bell correlations and nonlocality, as well as the absence of signalling.\footnote{Certainly it provides an appealing straightforward framework for considering the regime in which Initial Control is absent -- we simply need to turn off PH.} 
    \subsection{Fruitful interactions}
     Finally, WS express the hope that the interaction between quantum physics and the causal modelling literature will be fruitful in both directions. Like them, we are relying heavily on that literature, in our case for the notions of a collider and selection bias. We note that if our proposal turns out to be correct, it offers a substantial plum in the other direction, from physics to causal modelling. If correct, our proposal implies that explanation of one of the central features of the quantum world turns out to require the notion of a \textit{constrained} collider. This notion is not presently on the menu in the causal modelling literature, so far as we know.

\end{document}